\documentclass{aa}  
\usepackage{graphicx}
\usepackage{txfonts}
\usepackage{natbib}
\usepackage{booktabs}
\usepackage{tabularx} 
\bibpunct{(}{)}{;}{a}{}{,}   
\begin{document}
   \title{Unveiling the nature of six HMXBs through IR spectroscopy\thanks{Based on observations collected at the European Southern Observatory, Chile (Programme IDs: 077.D-0568 and 079.D-0562)}}


   \author{E. Nespoli
          \inst{1}
          \and
          J. Fabregat\inst{1}
          \and
          R. E. Mennickent\inst{2}
          }

   \offprints{Elisa Nespoli}

   \institute{Observatorio Astron\'omico de la Universidad de Valencia, Edifici Instituts d'Investigaci\'o. Pol\`igon La Coma, Paterna, Valencia, Spain\\
       \email{elisa.nespoli@uv.es}
   \and Departamento de F\'isica, Universidad de Concepci\'on, Casilla 160-C, Concepci\'on, Chile
          }
              

\date{Received February 25, 2008; accepted April 26, 2008}

 
  \abstract
   {The International Gamma-Ray Astrophyiscs Laboratory (INTEGRAL) is discovering a large number of new hard X-ray sources, many of them being HMXBs. The identification and spectral characterization of their optical/infrared counterparts is a necessary step to undertake detailed study of these systems. In particular, the determination of the spectral type is crucial in the case of the new class of Supergiant Fast X-ray Transients (SFXTs), which show X-ray properties common to other objects.}
  {Our goal is to perform spectral analysis and classification of proposed counterparts to HMXBs in order to characterize the system they belong to.}
   {We used the ESO/NTT SofI spectrograph to observe proposed IR counterparts to HMXBs, obtaining $K_{s}$ medium resolution spectra ($R = 1320$) with a S/N $\gtrsim$ 100. We classified them through comparison with published atlases.}
  {We were able to spectrally classify the six sources. This allowed us to ascribe one of them to the new class of SFXTs and confirm the membership of two sources to this class. We confirmed the spectral classification, derived from optical spectroscopy, of a known system, 4U 1907-09, showing for the first time its infrared spectrum. The spectral classification was also used to estimate the distance of the sources. We compared the extinction as derived from X-ray data with effective interstellar extinction obtained from our data, discussing the absorption component due to the circumstellar environment, which we observed in four systems; in particular, intrinsic absorption seems to emerge as a typical feature of the entire class of SFXTs. }
   {}


   \maketitle
%

\section{Introduction}

 \begin{table*}[!ht]
      \caption[]{NTT journal of observations. We report in the fourth column the net accumulated exposure time. Column five gives the obtained signal-to-noise ratio. The references list in the last column relates to the identification of the optical/infrared counterpart.}
         \label{table:logobs}
        \centering
            \begin{tabular}{lcccccl}
             \hline
             \hline
             \noalign{\smallskip}
              Source  & K mag &  Start time (UT) &  Exp. time (s) & S/N & IR Counterpart & Reference\\
             \noalign{\smallskip}
              \hline
             \noalign{\smallskip}
            IGR J16207--5129 & 9.1  & 2006-07-14 23:19   & 600 &  100 &  2MASS J16204627-5130060& Tomsick et al. (2006)\\
            IGR J16465--4507 & 9.8 & 2007-05-26 05:03 & 240  & 100 &  2MASS J16463526-4507045 & Zurita and Walter (2004)\\
            IGR J16479--4514 & 9.8 & 2007-05-26 05:11 & 240   & 100 & 2MASS J16480656-4512068 & Kennea et al. (2005)\\
            AX J1841.0--0536 & 8.9 & 2006-07-14 03:57 & 600  & 180 &  2MASS J18410043-0535465& Halpern et al. (2004)\\
            4U 1907+097 & 8.8 & 2006-07-15 07:37 & 600 & 130 & 2MASS J19093804+0949473 & Schwartz et al. (1980)\\            
            IGR J19140+0951 & 7.1 & 2006-07-14 04:46 & 360 & 130  & 2MASS J19140422+0952577 &  in't Zand et al. (2004)\\
            \noalign{\smallskip}
            \hline
         \end{tabular}
  \end{table*}
  
High Mass X-Ray Binaries (HMXBs) are systems composed of an early-type massive star and an accreting compact object. All sub-groups of HMXBs involve OB type stars and are commonly found in the galactic plane and in the Magellanic Clouds, among their OB progenitors.\

The majority of the known systems are Be/X-ray Binaries (BeXRBs), consisting of a neutron star accreting matter from the circumstellar equatorial disc of a Be star. Most of them are transient, exhibiting short and bright outbursts ($L_{X}\sim 10^{36} - 10^{37}$ erg s$^{-1}$ in the case of Type I outbursts, generally close to the periastron passage of the neutron star; $L_{X} \ge 10^{37}$ erg s$^{-1}$ in the case of Type II outbursts).

In the second major class of HMXBs, the Supergiant X-ray Binaries (SGXRBs), the counterpart is an early supergiant star, feeding the compact object with its radially outflowing stellar wind. As a result, the SGXRBs are, generally, persistent systems ($L_{X}\sim 10^{36}$ erg s$^{-1}$). \\

The five-year INTEGRAL data possibly reveal a different scenario. A recent subgroup has been proposed by \citet{negue06}, named Supergiant Fast X-ray Transients (SFXTs): these objects, associated with a supergiant companion, occasionally undergo a short period of X-ray activity lasting less then a day, typically a few hours \citep{sguera05}, with a very different behavior from those observed in other X-ray binaries. These outbursts show very sharp rises, reaching the peak of the flare in $\lesssim$1 hour. The decay is generally of a complex kind, with two or three further flares. The physical reason for these fast outbursts is still unknown, although theoretical speculations would connect them to some sort of discrete mass ejection from the supergiant donor \citep{gol03} or to wind variability \citep{int05}, or to the possible presence of a second, equatorial, wind component \citep{sid07} . Due to high interstellar absorption and to the transient nature of these sources, SFXTs are difficult to detect, and in most cases, the sources had not been detected by previous missions. To date, six objects have been firmly characterized as SFXTs, but many other systems are likely candidates, and their number has grown rapidly since the launch of INTEGRAL \citep{wink03}, so that they could actually constitute a major class of X-ray binaries.

Up to now, the INTEGRAL survey of the Galactic Plane and central regions has revealed the existence of more than 200 sources \citep{bird07,bodag07} in the energy range 20--100 keV, with a position accuracy of $2'-3'$, depending on count rate, position in the FOV and exposure. A large fraction of the newly discovered sources are found to be heavily obscured, displaying much larger column densities ($N_{H}\gtrsim 10^{23}$ cm$^{-2}$) than would be expected along the line of sight \citep[see][]{kuulkers05}. These sources were missed by previous high-energy missions, whose onboard instruments were sensitive to a softer energy range. Moreover, optical counterparts to these obscured sources are poorly observable due to the high interstellar extinction, with $A_{V}$ in excess of up to $\sim20$ mag.\\

 In this context, the recent availability of infrared spectroscopy has emerged as a strong tool to characterize these systems and, together with high-energy data, reveal the HMXB sub-class they belong to. This results in the identification of the mass transfer process of the system, with information about the intrinsic physics of the X-ray binary. The need for low energy data is particularly urgent in the case of SFXTs, which show X-ray properties common to other objects (such as RS CVs binaries and Low Mass X-ray binaries) and thus crucially require the spectral classification of their counterpart in order to be properly discerned.\\


In this paper we present spectral analysis and classification of six HMXBs discovered (or re-descovered) by INTEGRAL. The selected IGR sources are the following: IGR J16207--5129, IGR J16465--4507, IGR J16479--4514, AX J1841.0--0536 and IGR J19140+0951. We also included the well known system 4U 1907+09 since the spectral classification of its counterpart has been a matter of debate in the past, and no infrared spectra have been published up to now. The first three sources are located in the direction of the Norma-arm tangent region, the fourth in the Scutum-arm tangent region, the fifth and the last one in the Sagittarius arm tangent. In the next section we describe the observations and data reduction; in Section \ref{results} we report the obtained spectra, analyze their features and propose a classification; we calculate the interstellar hydrogen column density and estimate the distance to each source; in section \ref{discussion} we discuss our results, before concluding.  

Preliminary results of our data analysis for AX J1841.0--0536 and IGR J19140+0951 were published in \citet{nespoli07}.


\section{Observations and data analysis}    \label{observations}

We selected proposed counterparts, choosing sources observed by X-ray missions like XMM or Chandra, which can produce a very small error circle and facilitate the detection of the counterpart. 

Data were obtained in visiting mode during two observing runs, in July 2006 and May 2007 respectively, at the European Southern Observatory (ESO). The employed instrument was the SofI spectrograph \citep{sofi98}, on the 3.5m New Technology Telescope (NTT) at La Silla, Chile. Table \ref{table:logobs} reports the observation log, including, for each spectrum, the retrieved signal-to-noise ratio (S/N).

We used the long slit spectroscopy mode, at medium resolution ($R = 1320$) with a $K_{s}$ grism  and  1\arcsec\ width slit. The instrument large field objective provided a FOV of $4.92'  \  \times \ 4.92'$. The sky had thin cirri on 2006 July 14th and 15th, while it was generally clear on 2007 May 26th. Seeing averaged between $0.9''$ and $1.2''$, with the exception of the observation of IGR J16749--4514 which was performed with a seeing of $1.6''$.\\

In order to ensure accurate removal of atmospheric features from the spectra, we followed a strategy similar to that outlined by \citet{clark2000}. At the telescope, we observed an \mbox{A0 - A3} \mbox{III-V} standard star immediately before or after each target and a G2-3 V star once per hour in order to obtain very small differences in airmass (differences between 0.01 and 0.04 airmasses were accomplished). To compute the telluric features in the region of the \ion{H}{i} 21\,661 \AA\ (Brackett-$\gamma$ line, or Br$\gamma$), which is the only non-telluric feature in the A-star spectra, we employed the observed G-star spectra divided by the solar spectrum\footnote{We used the NSO/Kitt Peak FTS solar spectrum, produced by NSF/NOAO} properly degraded in resolution. The dispersion solution obtained for the SofI spectra was also applied and the spectra of the A star, G star and the solar one were aligned in wavelength space. A telluric spectrum for each scientific target was obtained by patching into the A-star spectrum the ratio between the G star and the solar spectrum in the Br$\gamma$ region (we selected the range 21\,590 - 21\,739 \AA). 

Typical integration times for standard stars were between 3 and 7 minutes.\\

Data reduction was performed using the IRAF\footnote{IRAF is distributed by the National Optical Astronomy Observatories which is operated by the Association of Universities for Research in Astronomy, Inc. under contract with the National Science Foundation} package, following the standard procedure. We first corrected for the inter-quadrant row cross-talk, a feature that affects the SofI detector; we then applied sky subtraction; we employed dome flat-fields and extracted one dimensional spectra. Wavelength calibration was accomplished using Xenon and Neon lamp spectra. Spurious features, such as cosmic rays or bad pixels, were removed by interpolation, when necessary. The reduced spectra were normalized by dividing them by a fitted polynomial continuum. We corrected for telluric absorption, dividing each scientific spectrum by its corresponding telluric spectrum, obtained as described above. A scale and a shift factor were applied to the telluric spectrum, to best correct for the airmass difference and the possible wavelength shift; the optimum values for these parameters were obtained using an iterative procedure that minimizes the residual noise. 


\section{Results}    \label{results}

In this section we present the results of spectral classification and analysis for each target. The field of NIR spectral classification is still very young and the level of required S/N and resolution to perform a quantitative profile analysis are very high ($R \sim 12\,000$ and S/N $\gtrsim$ 250), especially for young massive stars, as shown by \citet{hanson05-2}. The difficulty of the analysis depends on the few lines available and the relatively large uncertainty in their strength, due to their intrinsic weakness; moreover, some significant spectral regions, specifically through the 20\,580 He I and the Br$\gamma$ features, pose systematic complications because of the strong telluric absorption. 

Under this premise, our analysis will be qualitative, based on the comparison with available NIR spectral atlases \citep{hanson96,hanson05}. According to our estimation, this approach implies that the resulting spectral classification is provided with an uncertainty of no more than one luminosity subclass. The spectral type is precise up to one subtype.

Some of the identified features exhibit a 10-20 \AA\ displacement with respect to the nominal values, consistent with the instrumental resolution. Greater displacements, up to a maximum of 29 \AA, are found corresponding to lines characterized by a complex profile, such as Br$\gamma$, or placed in a region of strong telluric absorption, such as both \ion{H}{i} 20\,581 \AA\ and Br$\gamma$. 

In the next section we will classify the spectra, showing the features we were able to identify, with the corresponding equivalent width. Although it was pointed out \citep{hanson96} that equivalent widths may show variations between stars, we report them for completeness. The values we calculated are affected by an average 15$\%$ error.
For each object, a figure displays the spectrum we obtained, together with some comparative spectra from the atlas of \citet{hanson96} in a separate box. Table \ref{lines} summarizes the line identifications for all the observed targets.

\subsection{Spectral analysis and classification}

\subsubsection{IGR J16207--5129}
 The source was discovered by IBIS/ISGRI in the first Galactic Plane Survey performed by INTEGRAL \citep{bird04}, which measured a flux of 3.8 $\pm$ 0.3 mCrab in the 20--40 keV range. Instead, only the upper limit of $<4$ mCrab was obtained in the 40--100 keV band.
Subsequent Chandra observations allowed the identification of the optical/infrared counterpart. This was associated with USNO-B1.0 0384-0560857 = 2MASS J16204627-5130060 by \citet{tom06}. The power-law spectral fit provided $N_{H} = 3.7^{+1.4}_{-1.2} \times 10^{22}$ cm$^{-2}$ and photon index equal to $0.5^{+0.6}_{-0.5}$, indicating an intrinsically hard source. The lower limit to the stellar temperature was estimated to \mbox{$>$ 18\,000 K}, revealing the presence of a very hot, massive star. From the fit of optical/IR spectral energy distributions, the distance was estimated to be between 3-10 kpc (3-9 kpc in the case of a supergiant classification). Subsequent optical spectroscopy from \citet{mas06} refined the distance estimate to 4.6 kpc.

 \citet{negue07}, from optical observations, constrained the spectral type to earlier than B1.\\

\begin{figure}
   \centering
      \includegraphics[width=9cm]{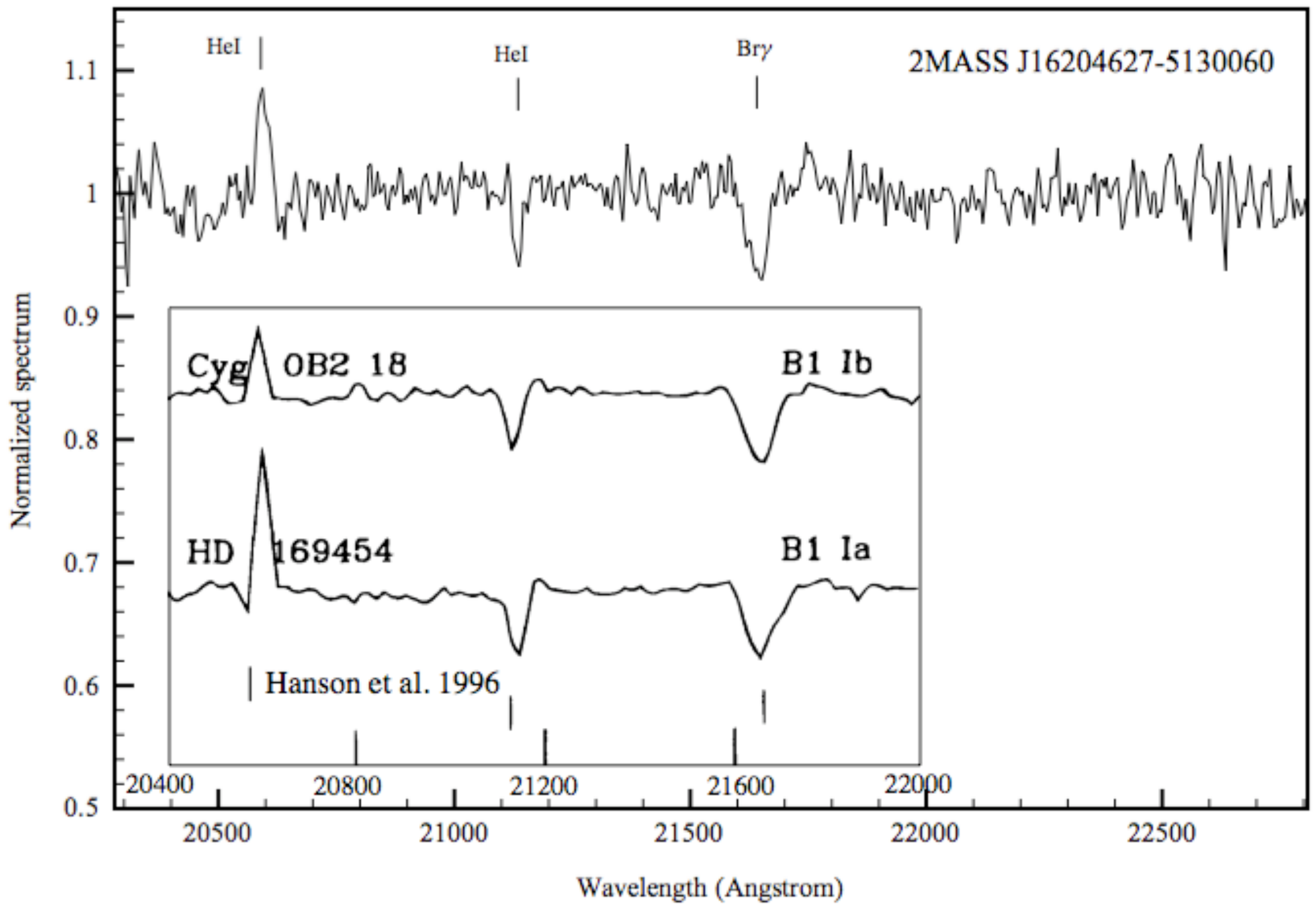}
      \caption{K$_{s}$ spectrum for 2MASS 16204627--5130060, the infrared counterpart of IGR J16207-5129. The positions of identified spectral features are marked by solid lines.
              }
         \label{fig:IGRJ16207}
   \end{figure}

Figure \ref{fig:IGRJ16207} shows the $K_{s}$ spectrum we obtained, with identified spectral features marked. \\


The spectrum shows no metal lines (no \ion{N}{iii} or \ion{C}{iv}), strong \ion{He}{i} 20\,581 \AA\ emission, \ion{He}{i} absorption at 21\,126 \AA\ and moderately strong Br$\gamma$ absorption. The atomic transitions observed are the typical of OB star spectra. The \ion{He}{i} 20\,581 \AA\ line is a prominent feature in supergiant stars, so that it is considered an important tracer of stars with extended atmospheres. It becomes weak or even disappears in main sequence stars and it is observed in emission in B type supergiants, whereas it is in absorption in O type supergiants \citep{hanson96}. The \ion{He}{i} 21\,126 \AA\ line is present in late O -- early B spectra.\\

For what was outlined above, and by comparing the relative strengths of the identified lines with those of \citet{hanson96,hanson05}, we estimate the spectral type of the IGR J16207--5129 counterpart to be B1 Ia. This allows us to classify the system as a SGXRB, as also inferred by \citet{negue07}.  

\subsubsection{IGR J16465-4507}   \label{IGRJ16465}

The source was discovered by INTEGRAL during its only observed X-ray flare on 2004 September 6th-7th \citep{lut04}.
Using data of XMM/Newton, \citet{lut05} found pulsations with period $P_{s}$ = 228 s in the X-ray flux and high photoabsorption, with $N_{H} \sim 7 \times 10^{23}$ cm$^{-2}$. The single star falling in the XMM/Newton error circle was identified by \citet{zur04} as the counterpart to the X-ray source, and associated with 2MASS J16463526-4507045. 

From an optical spectrum, \citet{negue06} classified the source as a B1 Ib supergiant, subsequently refining their classification to B0.5 Ib \citep{negue07-a}. They also estimated the distance of the source to 8 kpc. 

Our IR data suggest a different spectral classification.\\

\begin{figure}
   \centering
      \includegraphics[width=9cm]{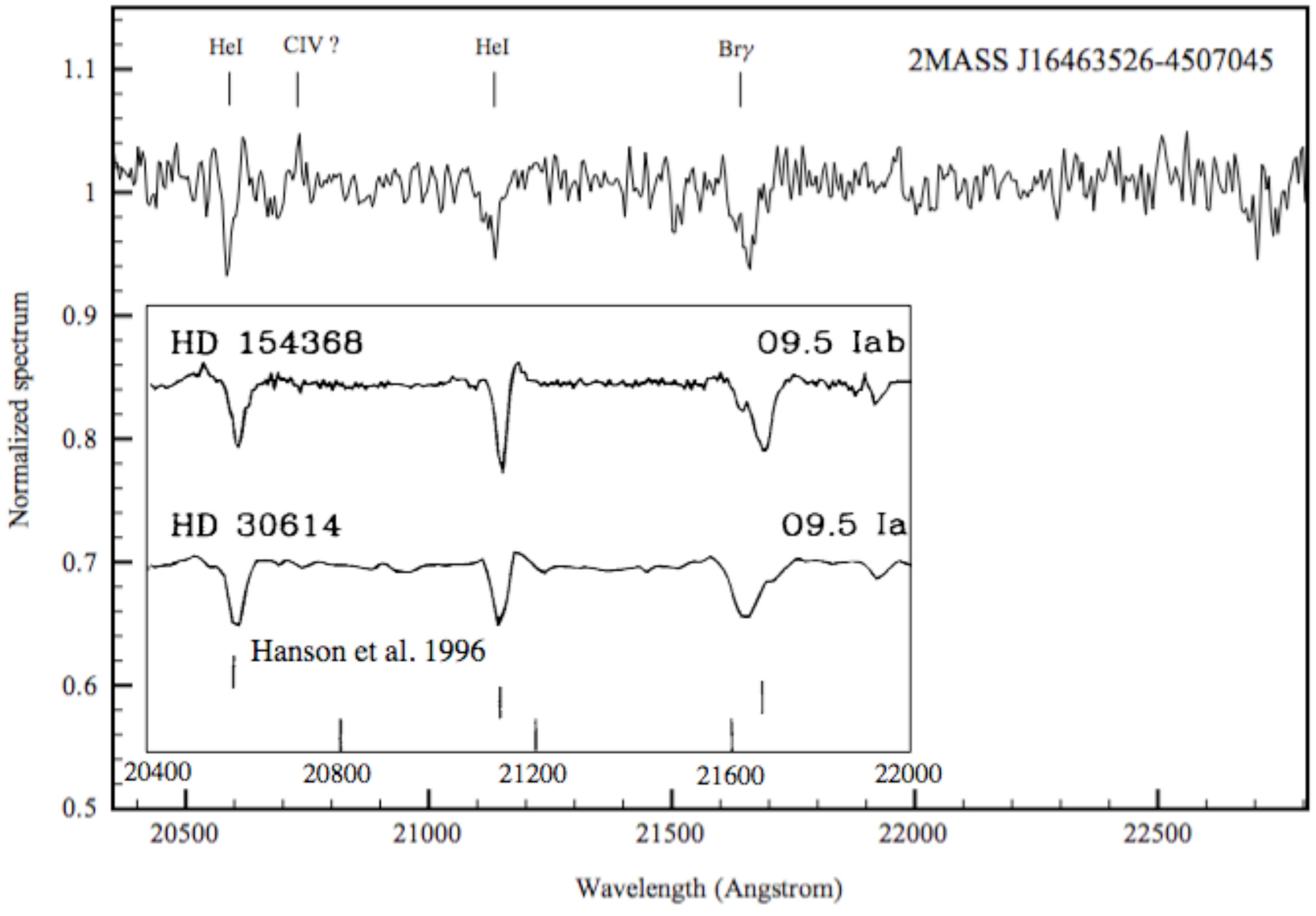}
      \caption{K$_{s}$ spectrum for 2MASS J16463526-4507045, the infrared counterpart of IGR J16465-4507. The positions of identified spectral features are marked by solid lines.
              }
         \label{fig:IGRJ16465}
   \end{figure}

Figure \ref{fig:IGRJ16465} shows the $K_{s}$ spectrum we obtained, with identified spectral features marked. \\ 

The spectrum shows \ion{He}{i} 20\,581 \AA\ in quite strong absorption; a feature at 20\,730 \AA\ is possibly identifiable with a faint blend of the \ion{C}{iv} transitions at 20\,690, 20\,780 and 20\,830 \AA, which appear in the atlas from \citet{hanson96} only in the very high S/N spectra. We observe \ion{He}{i} 21\,126 and Br$\gamma$, both in strong absorption.

We agree with \citet{negue07-a} in the classification of the counterpart as an early supergiant. However, we suggest that the observed features point to an earlier type than B1: in fact, the \ion{He}{i} 20\,581 \AA\ line is seen in emission in B1 supergiants, while it is in absorption in late-O supergiant. Comparing the relative strengths of the outlined features with those from the atlases of \citet{hanson96,hanson05}, we refine the spectral classification of the source to O9.5 Ia. We thus confirm the supergiant nature of the companion, which, together with the X-ray behavior of the system, classifies it as an SFXT, as proposed by \citet{negue06}.

\subsubsection{IGR J16479--4514}

The source was discovered by INTEGRAL \citep{molkov03} during an outburst. 
The X-ray spectrum is fitted with a power law with a high-energy cut-off, with spectral index $\Gamma =1.4$, and the column density is $N_{H} = 12\times 10^{22}$ cm$^{-2}$ \citep{lut05}.

IGR J16479--4514 showed short outbursts with very fast rises, observed in 2003 by \citet{sguera05}. Its X-ray behavior is thus typical of SFXTs \citep{negue06}, however the lack of an optical/infrared spectral classification prevented the possibility of enrolling it in this new class of objects.\\

The counterpart to the source was identified by \citet{ken05} through SWIFT observations.\\

\begin{figure}
   \centering
      \includegraphics[width=9cm]{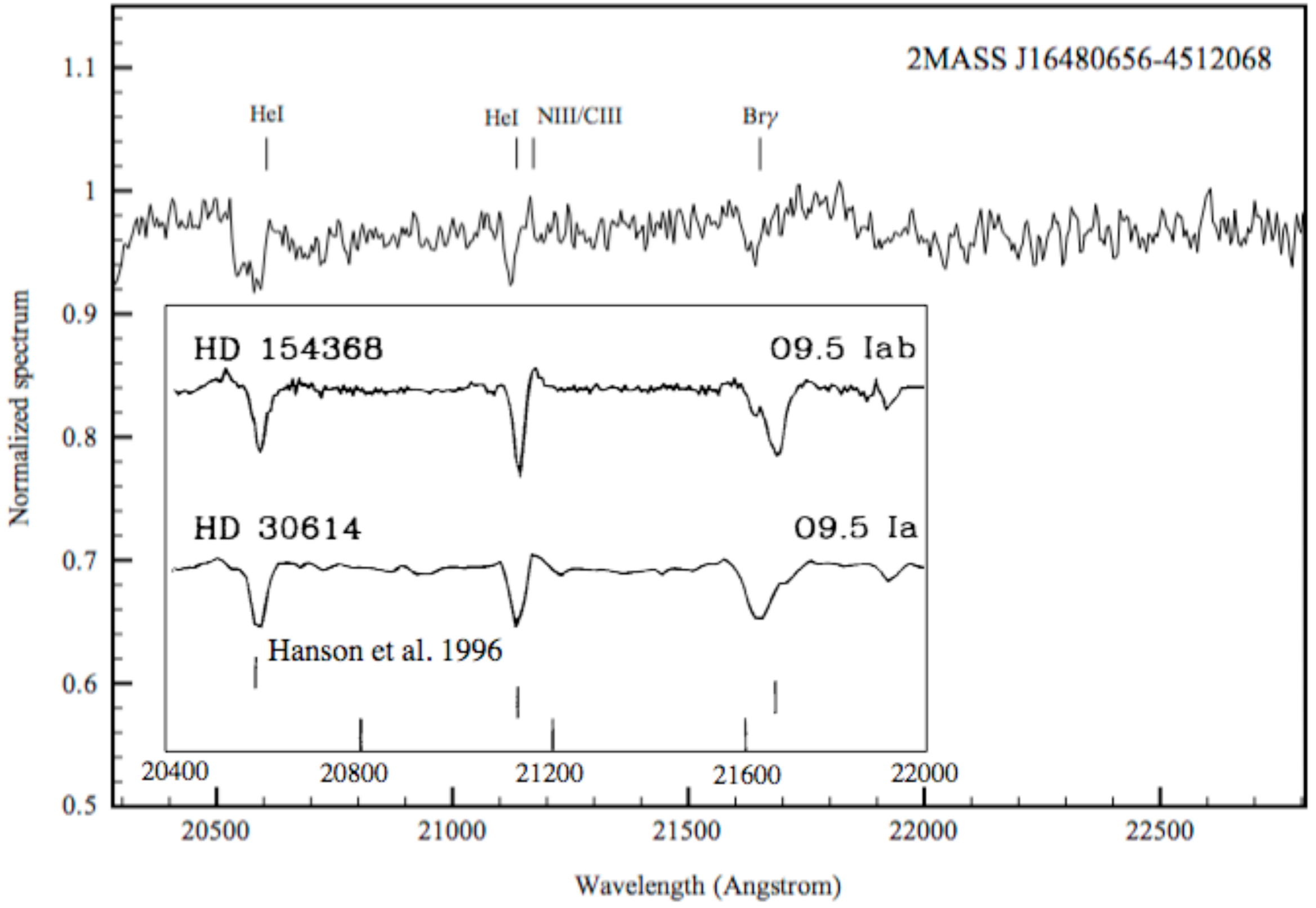}
      \caption{K$_{s}$ spectrum for 2MASS J16480656-4512068, the infrared counterpart of IGR J16479--4514. The positions of identified spectral features are marked by solid lines.
              }
         \label{fig:IGRJ16479}
   \end{figure}

Figure \ref{fig:IGRJ16479} shows the $K_{s}$ spectrum we obtained, with identified spectral features marked. 

The spectrum shows \ion{He}{i} 20\,581 \AA\ in strong absorption, well recognizable although affected by a clear telluric residual. The difference we obtained in airmasses between the telluric standard star and the target was in this case very low, equal to 0.004: we thus suppose that the poor correction of telluric absorption is due to the passage of a cirrus during the observation of the telluric standard.

 We can also detect absorption at \ion{He}{i} 21\,126, a weak \ion{N}{iii} 21\,155 \AA\ emission (which, according to \citet{hanson05}, could alternatively be \ion{C}{iii}), and moderately strong Br$\gamma$ absorption.\\

We conclude that the spectrum shows the typical features of a late O supergiant, especially the presence of  \ion{He}{i} 21\,126 (typical of late-O and early-B stars) in combination with \ion{He}{i} \mbox{20\,581 \AA,} present in supergiant stars, and seen in absorption in late-O supergiants. Through the comparison with the atlases from \citet{hanson96,hanson05}, we estimate the spectral type to be O9.5 Iab. This result, together with those from X-ray data, allows us to affirm that the object belongs to the new class of SFXTs.

\subsubsection{AX J1841.0--0536 }
   \begin{figure}[]
   \centering
      \includegraphics[width=9cm]{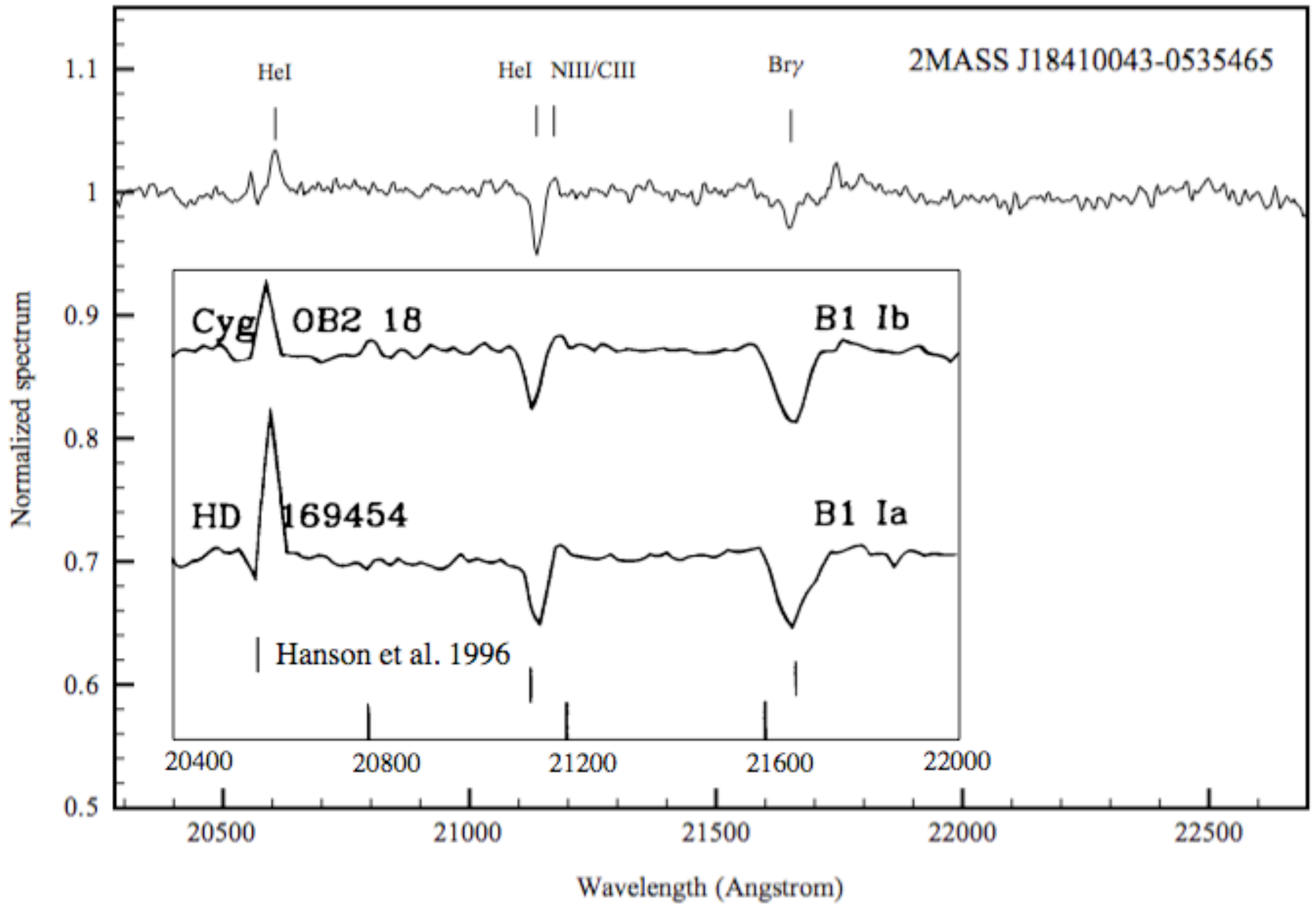}
      \caption{K$_{s}$ spectrum for 2MASS 18410043-0535465, the infrared counterpart of AX J1841.0-0536. The positions of identified spectral features are marked by solid lines.
              }
         \label{fig:AXJ1841}
   \end{figure}

AX J1841.0--0536 was discovered as a violently variable transient by ASCA in April 1994 \citep{bamba01}. The source showed multi-peaked flares with a sharp rise. Analysis of the ASCA data revealed that the source is a pulsar with \mbox{$P_{spin}$ = 4.7 s}. The spectral fit provided a value of $N_{H} = 3.2 \times 10^{22}$ cm$^{-2}$ \citep{bamba03}.  A fast outburst observed by INTEGRAL was attributed by \citet{halgot04} to this source.\\

A Chandra observation of the field revealed the counterpart to be 2MASS 18410043-0535465, a reddened star with weak H$\alpha$ in emission \citep{halpern04}, suggesting it was a Be star. 
 \citet{negue06}, from optical spectroscopy, proposed the star is instead a luminous B0-1 type, although with some uncertainty, classifying the system as an SFXT.\\
 
Figure \ref{fig:AXJ1841} shows the $K_{s}$ spectrum we obtained, with identified spectral features marked.

The spectrum shows \ion{He}{i} 20\,581 \AA\ emission, accompanied by a spurious feature, possibly due to poor telluric component removal; we observe absorption at \ion{He}{i} 21\,126, a weak \ion{N}{iii}  (\ion{C}{iii}) 21\,155 \AA\ emission; moreover, there is moderately strong Br$\gamma$ absorption. The side features of the Br$\gamma$ absorption profile are probably due to poor telluric correction, but they do not prevent us from measuring the equivalent width.\\ 

The observed transitions are typical of an early supergiant, and by comparison with the atlases from \citet{hanson96,hanson05}, we can conclude that the star is of B1 Ib type. Together with X-ray properties, this NIR spectral classification allows us to confirm the nature of the system as an SFXT.

\subsubsection{4U 1907+09}

   \begin{figure}[]
   \centering
      \includegraphics[width=9cm]{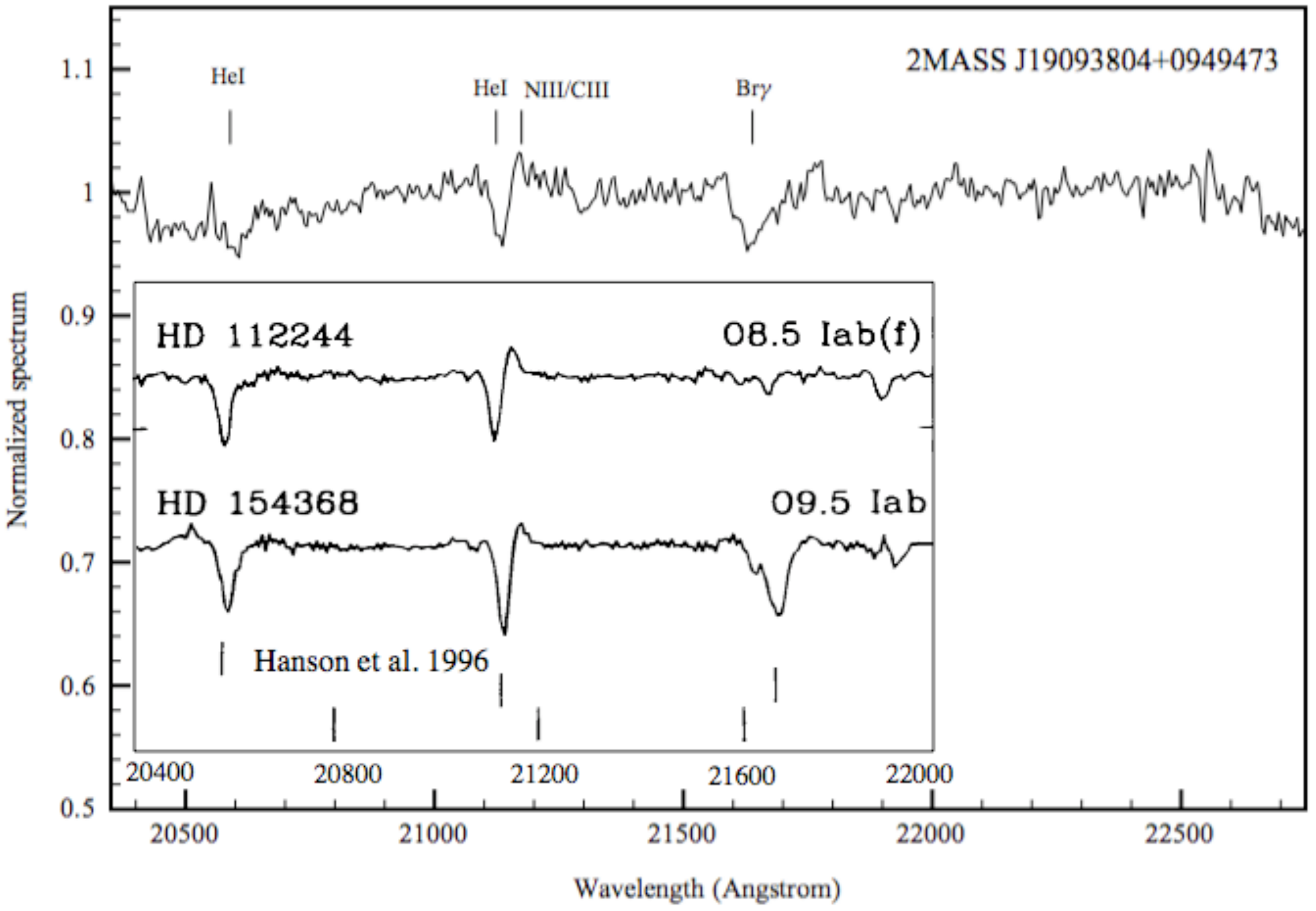}
      \caption{K$_{s}$ spectrum for 2MASS J19093804+0949473, the infrared counterpart of 4U 1907+09. The positions of identified spectral features are marked by solid lines.
              }
         \label{fig:4U1907}
   \end{figure}

The wind-accreting system 4U 1907+09 \citep{giacc71} is a known HMXB consisting of a neutron star in an eccentric ($e$ = 0.28) 8.3753 day orbit around its companion, which has been optically identified as a highly reddened star \citep{schwa80}. The spectral classification of the counterpart to 4U 1907+09 has been matter of debate. The presence of X-ray flaring seen twice per neutron star orbit \citep{mar80} had led some authors \citep[e.g.][]{maki84,copa87,Iye86} to the hypothesis of a Be star companion. However, this classification would require a distance of $<$1.5 kpc, which is in contradiction to the significant interstellar extinction measured in optical observations by \citet{vanker89}, who also classified the counterpart as a B supergiant. 
Using interstellar atomic lines of Na I and K I, \citet{cox05} set a lower limit of 5 kpc for the distance and proposed that the stellar companion is instead a O8-O9 Ia supergiant with an effective temperature of 30\,500 K, a radius of 26 $R_{\sun}$, a luminosity of $5 \times 10^5$ L$_{\sun}$,  and a mass loss rate of $7 \times 10^{-6}$ M$_{\sun}$ yr$^{-1}$.

Similarly to other accreting neutron stars, the X-ray continuum of 4U 1907+09 can be described by a power-law spectrum with an exponential turnover at 13 keV. The spectrum is modified by strong photoelectric absorption with a column density $N_{H} = 1.5 - 5.7 \times 10^{22}$ cm$^{-2}$ \citep[e.g.][]{copa87}.

We show for the first time an infrared spectrum of the source, which permits us to confirm the spectral classification as estimated from optical data.

Figure \ref{fig:4U1907} presents the $K_{s}$ spectrum we obtained, with identified spectral features marked. 

The spectrum shows \ion{He}{i} absorption both at 20\,580 \AA\ and at 21\,126 \AA, a weak N III (or C III) emission line at 21\,155 \AA\ and  strong Br$\gamma$ absorption (EW $<$ 4 \AA), the typical features of an early supergiant. The presence of \ion{He}{i} 20\,580 in absorption strongly constrains the spectral type to a late O star. By comparison with the atlases from \citet{hanson96,hanson05}, we conclude that the star is an O9.5 Iab. We thus confirm and refine the previous spectral classification.

\subsubsection{IGR J19140+0951}


   \begin{figure}
   \centering
      \includegraphics[width=9cm]{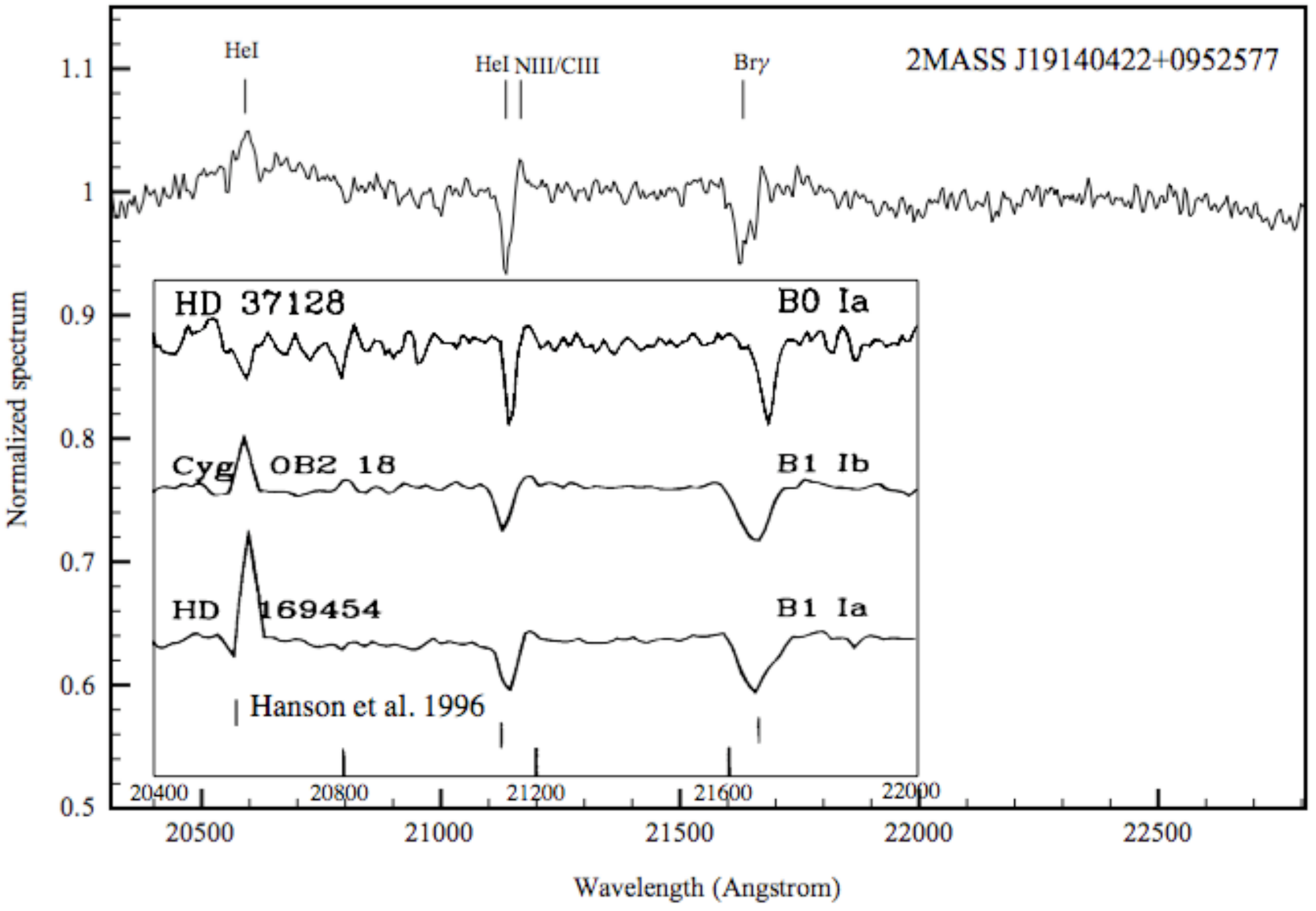}
      \caption{K$_{s}$ spectrum for 2MASS J19140422+0952577, the infrared counterpart of IGR J19140+0591. The positions of identified spectral features are marked by solid lines.
              }
         \label{fig:IGRJ19140}
   \end{figure}



\begin{table*}[!ht]
\caption{\small{$K$-band line identifications and spectral classification: for each target spectrum, the observed wavelength, the corresponding EW and our proposed classification are reported. Positive EWs refer to features in absorption.}}
\begin{center}
\resizebox{18cm}{!}{
\begin{tabular}{@{}lccccccccccl@{}} \toprule \toprule
   & \multicolumn{2}{c}{\ion{He}{i} (2s$^1$S - 2p$^1$P$^0$)}  & \multicolumn{2}{c}{\ion{C}{iv} (3p$^2$P$^0$ - 3d$^2$D)}  & \multicolumn{2}{c}{\ion{He}{i} (3p$^3$P$^0$ - 4s$^3$S), 21\,120 \AA } & \multicolumn{2}{c}{\ion{N}{iii} (\ion{C}{iii} )} & \multicolumn{2}{c}{Br$\gamma$}  \\ 
 & \multicolumn{2}{c}{20\,581 \AA} & \multicolumn{2}{c}{ 20\,690, 20\,780, 20\,830 \AA} & \multicolumn{2}{c}{+ HeI (3p$^1$P$^0$ - 4s$^1$S), 21\,130 \AA} & \multicolumn{2}{c}{21\,155 \AA} & \multicolumn{2}{c}{21\,661 \AA} & Spectral  \\
\cmidrule(r){2-3}
\cmidrule(r){4-5} 
\cmidrule(r){6-7}
\cmidrule(r){8-9}
\cmidrule(r){10-11}
 Object & Wavelength & EW & Wavelength & EW & Wavelength & EW & Wavelength & EW & Wavelength & EW  &  classification \\
             & [\AA] & [\AA]  & [\AA] & [\AA] & [\AA] & [\AA] & [\AA] & [\AA]  & [\AA] & [\AA] & \\ \midrule
IGR J16207--5129 & 20\,595  & -2.90 &      --       &     --       &  21\,134   & 1.23   &  --            &  --            &  21\,642  & 3.52 &  B1 Ia\\ 
IGR J16465--4507 &  20\,589 & 2.72  & 20\,730 & -0.42 &  21\,128      & 2.31   & --             & --             & 21\,654  &  3.25   & O9.5 Ia$^{*}$\\
IGR J16479--4514 & 20\,570  & 3.10  &     --         &     --        & 21\,123   & 1.23  & 21\,162   &  -0.39    & 21\,640  & 1.39  &  O9.5 Iab$^{*} $\\
AX J1841.0--0536 & 20\,599 & -0.94  & --             &    --          & 21\,135  &  1.64  & 21\,171  &  -0.23    &  21\,652 & 1.89   &  B1 Ib$^{*}$ \\
4U 1907+09            & 20\,602 & 1.12  &      --        & --              &  21\,130 &   1.84 &  21\,170 &  -0.25   & 21\,641   &  3.19  & O9.5 Iab\\
IGR J19140+0951 & 20\,591 & -1.38 &     --         & --              &  21\,136  &  1.56 &  21\,167  &  -0.33  &  21\,632 &   1.86 & B1 Iab 
   \\\bottomrule
\end{tabular}}
\begin{list}{}{}
\item[$^{\mathrm{*}}$] \small{Systems classified through this work as new or confirmed SFXTs.}
\end{list}
\end{center}
\label{lines}
\end{table*}


The INTEGRAL discovery of this source was reported by \citet{hann03}. Observations with the Rossi X-Ray Timing Explorer (RXTE) revealed a rather hard spectrum, fitted with a power law of photon index 1.6 and an absorption column density of $N_{H} = 6 \times 10^{22}$ cm$^{-2}$ \citep{swank03}. Timing analysis of the RXTE All-Sky Monitor (ASM) data showed an X-ray period of 13.55 days \citep{corbet04}.

\citet{hann04} presented high energy spectral analysis of the period of the discovery, concluding that the source manifests two distinct spectral behaviors, the first showing a thermal component in the soft X-ray and hard X-ray tail, the second being harder and possibly originating from thermal Comptonization. This second, low-luminosity, behavior was confirmed to be the preferred state of the source \citep{rodri05}.\\

The optical/infrared counterpart to IGR J19140+0951 was identified by \citet{int06}, from Chandra accurate position determination, as the heavily reddened \mbox{2MASS 19140422+0952577}.\\

Figure \ref{fig:IGRJ19140} shows the $K_{s}$ spectrum we obtained, with identified spectral features marked. 
The spectrum shows \ion{He}{i} \mbox{20\,581 \AA} emission, absorption at \ion{He}{i} 21\,126 \AA, a weak \ion{N}{iii} (or \ion{C}{iii}) emission feature at 21\,155 \AA, and moderately strong Br$\gamma$ absorption, typical features of an early supergiant. By comparison with the atlases from \citet{hanson96,hanson05}, we can conclude that the star is a B1 Iab type. Together with X-ray properties, this allows us to confirm the nature of the system as an SGXRB.\\
Our preliminary results were published in \citet{nespoli07}, and recently confirmed by  \citet{hann07}, who, from $K$- and $H$-band spectra, constrained the spectral type to a B0.5 supergiant. They also estimated the distance of the source as 5 kpc.

\subsection{Reddening and distance estimation}

From the identified spectral types, we obtained the intrinsic colors $(J-K)_{0}$ from \citet{weg94}; we then calculated, from 2MASS photometry, the instrumental colors $(J-K)_{2MASS}$, properly transformed through the formula from \citet{car01} \footnote{In its updated version at http://www.astro.caltech.edu/~jmc/2mass/ /v3/transformations/} to the \citet{bebe88} homogenized photometric system in order to estimate the infrared color excess $E(J-K)$. Assuming the mean extinction law ($R_{V}= 3.1$), from $A_{V}/E(J-K) = 5.82 \pm 0.10$ \citep{rieke85}, we obtained the total measured visual extinction $A_{V}$ and the corresponding hydrogen column density value from $N_{H}/A_{V} = 1.79 \pm 0.03 \times 10^{21}$ atoms cm$^{-2}$ mag \citep{pred95}. We were thus able to compare the retrieved interstellar value of $N_{H}$ with the one provided by X-ray data. In our calculation, we estimated errors through error propagation. Errors in the final values of $N_{H}$ are mainly due to errors in the infrared colors and of the transformation between the two photometric systems. 

We also estimated the distance of the six sources, applying the relation $M_{K} = K + 5 - 5$ log $d - A_{K}$.  For each source, $M_{V}$ was obtained from our proposed spectral type \citep{weg06}, the intrinsic color index $(K-V)_{0}$ from \citet{weg94} was used in order to calculate $M_{K}$, and the 2MASS $K$ magnitude was employed. We derived $A_{K}$ from the relation $A_{\lambda} / E(J-K) = 2.4 (\lambda)^{-1.75}$, for $ \lambda = 2.2\ \mu$m \citep{draine89}.\\

The results of our calculations are given in Table \ref{tab:reddening}, together with some crucial quantities used in the calculations or displayed for comparison. The distance estimation is mainly affected by the uncertainty in the value of the absolute magnitude $M_{V}$, which is due to two contributions, the errors given in the tabulated values of $M_{V}$ and the uncertainty in the spectral classification, from which the absolute magnitude is determined. The largest role is played by the errors in the mean tabulated values themselves (see Wegner 2006 for more details). The retrieved values for $d$ must thus be assumed with prudence.

 \begin{table*}[!ht]
      \caption[]{For each observed source in the first column, the obtained spectral classification, intrinsic infrared colors, 2MASS photometry, calculated infrared excess, hydrogen column density obtained from X-ray published measurements, effective interstellar column density obtained from our work and distance estimation are reported. See text for corresponding references.}
        \centering
         \label{tab:reddening}
            \begin{tabular}{lccccccc}
             \hline
             \hline
             \noalign{\smallskip}
              Source  & Spectral type &  $(J-K)_{0}$ &  $(J-K)_{2MASS}$ & $E(J-K)$& $N_{H}$ from X-ray data  & Interstellar $N_{H}$        & Distance\\
                               &                      &         [mag]       &       [mag]                 &   [mag]    &      [10$^{22}$ cm$^{-2}$]  &   [10$^{22}$ cm$^{-2}$]  & [kpc]\\             
               \noalign{\smallskip}
              \hline
             \noalign{\smallskip}
            IGR J16207--5129   &   B1 Ia         & -0.12   & 1.31    & 1.43  & $3.7^{+1.4}_{-1.2}$             &  $1.53 \pm 1.02$    & 6.1 \tiny{(-3.5, +8.9)}\\
            IGR J16465--4507   &   O9.5 Ia     & -0.15   & 0.69    & 0.84  & $72 \pm 6$           &  $0.87 \pm 0.56$    & 9.5 \tiny{(-5.7, +14.1)} \\
            IGR J16479--4514   &   O9.5 Iab   & -0.14   & 3.19   & 3.33   & $12 \pm 4$          & $ 3.47 \pm 2.16$     & 2.8  \tiny{(1.7, +4.9)} \\
            AX J1841.0--0536   &    B1 Ib         & -0.13   & 0.80   & 0.93  & 3.2             &  $ 0.97 \pm 0.64$    & 3.2  \tiny{(-1.5, +2.0)} \\
            4U 1907+097           &   O8.5 Iab   & -0.17    & 1.22  & 1.39   & 1.7 -- 5.7   & $ 1.45  \pm 0.87$    & 2.8  \tiny{(-1.8, +5.0)} \\            
            IGR J19140+0951   &   B0.5 Iab   & -0.12    & 1.50  & 1.62   &  $\sim6^{a}$            &  $ 1.68 \pm 1.5$    & 1.1  \tiny{(-0.8, +2.3)}  \\
            \noalign{\smallskip}
            \hline
         \end{tabular}
  \begin{list}{}{}
\item[$^{\mathrm{a}}$] A maximum value of $10.1\pm 0.2 \times10^{22}$ cm$^{-2}$ was reported by Rodriguez et al. 2005.
\end{list}
  \end{table*}

\section{Discussion}     \label{discussion}

Using near-infrared spectroscopy of six high-energy sources, IGR J16207--5129, IGR J16465--4507, IGR J16479--4514, AX J1841.0--0536, 4U 1907+097 and IGR J19140+0951, we classified their counterparts through comparison with published atlases. We found that all the observed systems have a supergiant companion. Our results, combined with information from X-ray data, is able for the first time to firmly include one source, IGR J16479--4514, in the newly discovered class of the SFXTs. Moreover, we can confirm with infrared data the identification of IGR J16465--4507 and AX J1841.0--0536 as SFXTs, as recently proposed by \citet{negue07-a} from optical spectra. \\

From our spectral classification, we estimated the distance of the six sources. The retrieved values are consistent with the location of the sources in the Norma (IGR J16207--5129, IGR J16465--4507, IGR J16479--4514), Scutum (AX J1841.0--0536) and Sagittarius arm (4U 1907+097 and IGR J19140+0951) regions respectively. This determination, although affected by some uncertainty,  can be considered an \emph{a posteriori} confirmation of the proposed spectral classification.\\

This work allowed us to calculate the extinction from IR data for the six systems. Recently, it has been pointed out \citep[see][]{kuulkers05,cha07} that INTEGRAL is revealing two new classes of supergiant HMXBs, the highly obscured HMXBs and the SFXTs. The first ones are characterized by strong intrinsic absorption, the second by strong and short X-ray outbursts. High, variable, hydrogen column densities have in some cases been measured for SFXTs (e.g. IGR J11215--5952: \citealt{smi06-a}; IGR J17391--3021: \citealt{smi06}), suggesting a possible intrinsic absorption for this class as well, and marking a potential overlap between the two new classes. The origin and position (around the compact object only, or enveloping the entire system) of the absorbing material are still a matter of debate, and only multiwavelength studies are able to address the problem, distinguishing between the absorption in X-ray and in the IR/optical bands. 

 We calculated the effective interstellar extinction $A_{V}$ and converted it into hydrogen column density $N_{H}$. Our results can be compared with the values obtained from X-ray data. If the two retrieved values are compatible within the corresponding errors, we face two possible scenarios: either the source of absorption is just the interstellar medium or there is a contribution from an extensive envelope around the whole binary system, if the extinction shows an excess of some orders of magnitude with respect to the estimated IS value. Conversely, if the reddening (measured from the IR colors) is low for the measured $N_{H}$ from X-rays, this reveals the presence of an additional source of extinction, which only affects the compact object in which the X-ray emission originates, and that can be regarded as the presence of absorbing material around it.
 
In our case, the comparison reveals that for four systems, IGR J16465--4507, IGR J16479--4514, AX J1841.0--0536 and IGR J19140+0951, the extinction measured at high energy shows an excess of one or two orders of magnitude with respect to that obtained from IR data,  probing that the material absorbing in the X-ray is concentrated around the neutron star. Our result is consistent with the detection of the Fe fluorescence line at 6.4 KeV in the X-ray spectra of all the four sources \citep{wal06,bamba01,rodri05}, which is considered a signature of the dense spherical envelope around the compact object.  

Usually, the so called highly absorbed IGR sources are identified from their measured $N_{H}$ being $\gtrsim 10^{23}$ cm$^{-2}$, i.e. one to two orders of magnitude higher than the assumed Galactic value of \mbox{$\sim10^{22}$ cm$^{-2}$} \citep{kuulkers05}. The comparison of X-ray data, which are sensitive to the absorption from the environment of the compact object, with infrared data, which are in general only absorbed by the interstellar medium, can be a strong and alternative criterion to recognize this class of highly absorbed sources.  \\
In particular, all the three identified SFXTs show strong intrinsic absorption from circumstellar material, reported as well for other members of this new class \citep[e.g.][]{negue06,pel06}, and so emerging as a common feature of the group. This scenario is relevant to discussions about the physical mechanism, still unknown, which powers the fast outbursts that these sources undergo. Together with their X-ray fast transient nature, the high intrinsic absorption can be regarded as one of the reasons why this class has remained undiscovered until only recently.  \\

A more complete, multiwavelength investigation is necessary to reveal the nature of these systems, especially if it is able to find and physically describe correlations between their X-ray behavior and IR properties. In the case of SFXTs this is possibly the only way to explore the physical mechanism driving the fast outbursts.

\section{Conclusions}

From near-infrared spectroscopy of the six HMXBs we have found that:

\begin{itemize}
\item[-] the proposed optical counterparts were confirmed and the spectral classification of the sources provided;
\item[-] one source, IGR J16479--4514, was added to the SFXTs and the confirmation of IGR J16465--4507 and AX J1841.0--0536 as members of the class was proven with infrared data;
\item[-] the comparison between $N_{H}$ obtained from X-ray data and interstellar extinction from our data showed for four systems (IGR J16465--4507, IGR J16479--4514, AX J1841.0--0536 and IGR J19140+0951) the presence of an absorbing envelope, strictly confined around the compact object;
\item[-] all the three identified SFXTs are intrinsically absorbed, suggesting that this might be a characteristic of the class;
\item[-] the distance estimation, compatible with the location of the sources in the respective galactic arms, is a possible confirmation of the spectral classification provided here. 
\end{itemize}

  \begin{acknowledgements}
The work of EN and JF is supported by the Spanish Ministerio de Educaci\'on y Ciencia, and FEDER, under contract AYA 2007-62487. EN aknowledges a ``V Segles'' research grant from The University of Valencia. RM acknowledges Fondecyt 1070705 and Chilean Center for Astrophysics FONDAP 15010003. 
  
      \end{acknowledgements}

\bibliographystyle{aa}
\bibliography{HMXB_IRspec_arxiv}

\end{document}